\documentclass[structabstract]{aa}  
\usepackage{natbib}
\bibpunct{(}{)}{;}{a}{}{,} 
\usepackage{graphicx}
\usepackage{txfonts}
\def\gsim{\hbox{\raise0.5ex\hbox{$>\lower1.06ex\hbox{$\kern-0.94em{\sim}$}$}}}
\begin{document}
   \title{Searching for pulsed emission from XTE~J0929--314 at high radio frequencies}

   \author{M. N. Iacolina
          \inst{1}
          \and
          M. Burgay\inst{2}
          \and
          L. Burderi\inst{1}
          \and
          A. Possenti\inst{2}
          \and
          T. Di Salvo\inst{3}
          }

   \institute{Universit\`a di Cagliari, Dipartimento di Fisica, 
              SP Monserrato-Sestu km 0.7, 09042 Monserrato (CA), Italy\\
              \email{iacolina@ca.astro.it}
         \and
             INAF-Osservatorio Astronomico di Cagliari,
             Loc. Poggio dei Pini, Strada 54, 09012 Capoterra (CA), Italy
         \and
             Universit\`a di Palermo, Dipartimento di Scienze Fisiche 
             ed Astronomiche, via Archirafi 36, 90123 Palermo, Italy\\
             }

   \date{Received ; accepted }

  \abstract 
{} 
{The aim of this work is to search for radio signals in the quiescent phase of accreting millisecond X-ray pulsars, in this way giving an
ultimate proof of the recycling model, thereby unambiguously establishing that accreting millisecond X-ray pulsars are the progenitors of radio
millisecond pulsars.}  
{To overcome the possible free-free absorption caused by matter surrounding accreting millisecond X-ray pulsars in their 
quiescence phase, we performed the observations at high frequencies. Making use of particularly precise orbital and spin parameters
obtained from X-ray observations, we carried out a deep search for radio-pulsed emission from the accreting millisecond X-ray pulsar 
XTE~J0929--314 in three steps, correcting for the effect of the dispersion due to the interstellar medium, eliminating the 
orbital motions effects, and finally folding the time series.}  
{No radio pulsation is present in the analyzed data down to a limit of 68 $\mu$Jy at 6.4 GHz and 26 $\mu$Jy at 8.5 GHz.}  
{We discuss several mechanisms that could prevent the detection, concluding that beaming factor and intrinsic low luminosity are the
most likely explanations.}


\titlerunning{Searching for pulsed radio emission from XTE~J0929--314}
\authorrunning{M. N. Iacolina et al.}
   \maketitle
%

\section{Introduction}\label{intro}

The neutron star low-mass X-ray binaries (NS-LMXB) are systems
containing a neutron star (NS) believed to have a weak magnetic field
(B $ < 10^{10}$ G) and accreting matter from a low-mass (M $ \leq
1$ M $_{\odot}$) companion.  A special subgroup of NS-LMXBs is that of the
neutron star soft X-ray transients (NS-SXT; see e.g. White et al.
1995\nocite{wnp95}).  These systems are usually found in a
quiescent state with luminosities in the range $10^{31}-10^{34}$ erg
$\cdot$ s$^{-1}$. On occasion they exhibit outbursts with peak
luminosities in the 0.5-10 keV range between $10^{36}$ and $10^{38}$
erg $\cdot$ s$^{-1}$, during which their behaviour closely resembles
that of the persistent NS-LMXBs (Campana et
al. 1998\nocite{ccm+98}). According to the recycling scenario (Alpar
et al. 1982\nocite{acrs82}; Bhattacharya \& van den Heuvel
1991\nocite{bv91}), the NS-LMXBs are the progenitors of the millisecond
radio pulsars (MSPs), reaccelerated by mass and angular momentum
transfer from the companion star.  

Although widely accepted, there is
no direct evidence to confirm this scenario yet. However, in 1998, the
idea that NSs in NS-LMXBs are spinning at millisecond periods was
spectacularly demonstrated by the discovery of coherent X-ray
pulsation at 2.5 ms in SAX J1808.4--3658, an NS-SXT with an orbital
period P$_{\rm orb} \sim$ 2 hr (Wijnands \& van der Klis
1998\nocite{wv98}; Chakrabarty \& Morgan 1998\nocite{cm98}).  For
almost four years, SAX J1808.4--3658 has been considered a rare
object in which some peculiarity of the system (e.g. its inclination)
allowed detection of the NS spin; however, the situation is now
dramatically changed as nine other NS-SXTs in outburst have been
discovered in which coherent X-ray pulsations in the millisecond range
have been found (e.g. Wijnands et al.\nocite{wij06} 2006; Krimm et
al.\nocite{kmd+07} 2007; Casella et al.\nocite{cap+08} 2008;
Altamirano et al.\nocite{acp+08} 2008).  We are, therefore, facing a
new class of astronomical objects (dubbed accreting millisecond X-ray
pulsars: AMXPs) that may constitute the bridge between
accretion-powered rapidly-rotating and rotation-powered NSs; in
particular, the detection of radio pulsations from these sources during 
quiescence would be the ultimate proof of the validity of the recycling model.

During the mass transfer, the plasma from the companion settles into
an accretion disk, whose inner rim is located at a distance r$_{\rm m}$ from
the NS centre, with r$_{\rm m}$ the so-called magnetospheric radius, which is
equal to a fraction $ \phi\leq 1$ of the Alfv\`en radius, at which the
ram pressure of the infalling matter from the accretion disk balances
the pressure of the NS magnetic field :
 \begin{equation}
\rm r_m=1.0\times 10^6 \ \phi\ \mu_{26}^{4/7}\ \dot{m}^{-2/7}_{E}m_1^{-1/7}\hspace{0.5cm} cm.
\end{equation}
Here $\mu_{26}$ is the magnetic moment in units of $10^{26}$ G $\cdot$
cm$^3$, R$_6$ is the NS radius in units of $10^6$ cm, $\rm \dot{m}_E$
the Eddington accretion rate in M$_{\odot}\cdot$ yr$^{-1}$ (for
R$_6=1$, $\rm \dot{m}_E=1.5 \times 10^{-8}$ M$_{\odot}\cdot$ yr$^{-1}$ and
scales with the radius of the compact object), and m$_1$ the NS mass
in solar masses.

For the radio emission mechanism to switch on in a rotating magnetized
NS, the space surrounding the NS must be free of matter up to the
light cylinder radius (at which the speed of the material co-rotating
with the NS would be equal to the speed of light):
 \begin{equation}
\rm r_{lc}=5\times 10^6\ P_{-3}\hspace{0.5cm} cm
\end{equation}
where P$_{-3}$ is the pulse period in milliseconds.

During an outburst, the occurrence, in some cases, of type I bursts
and the observation of coherent X-ray pulsation indicate that at the
origin of the observed luminosity is the accretion mechanism onto the
NS surface. As a consequence coherent radio emission cannot occur in
this phase. In the quiescent state, the 0.5-10 keV luminosity is
detected at levels ranging between $10^{31.5}$ and $10^{34}$ erg
$\cdot$ s$^{-1}$.  Accretion of matter at a lower rate was originally
proposed to account for this lower luminosity emission. However
detailed studies of the X-ray spectrum in quiescence (Rutledge et
al. 2001\nocite{rbb+01}) and of the thermal relaxation of the NS crust
during this phase (Colpi et al. 2001\nocite{cgpp01}) suggest that the
cooling of the periodically warmed up NS surface (Brown, Bildsten, \&
Rutledge 1998\nocite{bbr98}) is a viable explanation for the bulk of
the luminosity in quiescence.  If this is the correct interpretation,
there should be no mass accreted onto the neutron star surface
during the X-ray quiescent phase, hence $\rm \dot{m} \sim 0$, and
thus, plausibly,
\begin{equation}
\rm r_m \geq r_{lc}.
\end{equation}
The timescale for the expansion of the magnetospheric radius beyond
the light cylinder radius, in response to a sudden drop in the mass
transfer rate, is much shorter (Burderi et al. 2001) than the typical
duration (approximately years) of a phase of quiescence in an NS-SXT,
allowing the radio pulsar to switch on in principle.

\subsection{Absorption effect of the matter surrounding the system}\label{tff}
In 2001 Burderi et al. proposed a model able to explain the failure of
previous and subsequent searches (Burgay et
al. 2003\nocite{bbpd03} and references therein) for pulsed radio
emission from quiescent NS-SXTs.  As mentioned above, a temporary
significant reduction of the mass-transfer rate may cause the
magnetospheric radius to overcome r$_{\rm lc}$, thus allowing a radio
pulsar to switch on.  In some cases, even if the secular mass-transfer
rate is restored, the accretion of matter onto the NS can be inhibited
because the radiation pressure from the rotating dipole may be capable of
ejecting most of the matter overflowing from the
companion out of the system. This phase has been defined as {\it radio-ejection}. One of
the strongest predictions of this model is the presence, during the
radio-ejection phase (hence during X-ray quiescence), of a strong wind
of matter emanating from the companion star swept away by the
radiation pressure of the pulsar (see Di Salvo et al. 2008a\nocite{dbr+08a} for a discussion of a possible secular
evolution of these systems).  This matter, as well as that residual
from a previous outburst, can play a role in absorbing the radio
signal. Following Burderi et al. (2001\nocite{bpd+01}) and Burgay et
al. (2003), we can estimate the optical depth, $\rm \tau_{ff}$, at various
radio frequencies due to matter engulfing an NS-SXT in its quiescent
phase:
\begin{equation}\label{eq:tauff}
\rm \tau_{ff} = 1.6 \times 10^3 \times \frac{\gamma^2 \ \dot{m}_{-10}^2 (X+0.5Y)^2\ F(m_1,m_2)\ Ga}{m_1^{5/3}\ T^{3/2}_4\ P_h^{4/3}\ \nu_9^{2}\left(1+ m_1/m_2\right)^{5/3}}
\end{equation}
where $\rm \dot{m}_{-10}$ is the mass transfer rate in outburst in units
of $10^{-10}$~M$_{\odot}\cdot$ y$^{-1}$, X and Y are the mass fraction
of hydrogen and helium respectively, $\gamma$ is the fraction of
ionized hydrogen, T$_4$ the temperature of the outflowing matter in
units of $10^4$ K, $\nu_9$ the frequency of the radio emission in
units of $10^9$ Hz, P$_{\rm h}$ is the orbital period in hours, m$_1$ and m$_2$ are the masses of the NS and its companion in solar masses, 
F (m$_1$, m$_2$) = 1 $-$ 0.462 m$_2$ / (m$_1$ + m$_2$), Ga~=~Ga (T$_4, \nu_9$,
Z) = 1.00 + 0.48 ($\log $T$_4 - \log$ Z) $-$ 0.25 $\log \nu_9$ takes
into account the dependencies of the Gaunt factor from the
temperature, the atomic number Z, and the frequency.

One example of a system in the radio-ejection phase is PSR J1740--5340,
an eclipsing MSP, with a spin period of 3.65 ms located in the
globular cluster NGC 6397 (D'Amico et al. 2001). For this source
$\tau_{\rm ff} = 2.8$ at $\nu_9=1.4$ (with X = 0.7, Y = 0.3, $\gamma$ = 1,
T$_4=1$, Ga = 0.94 for Z = 1.1). This shows that the effect of
free-free absorption at 1.4 GHz is important, but not severe, for this
system. The radio signal from this PSR is indeed visible at this
frequency although it is eclipsed and distorted at many orbital phases
(D'Amico et al. 2001\nocite{dpm+01}).  X-ray millisecond pulsars are
very similar to PSR J1740--5340, the main difference being the orbital
period, which is $\sim$ 32 hr in the case of PSR J1740--5340 and typically ranges
from 40 min to $\sim$ 4 hr in the case of the X-ray millisecond
pulsars. The difference in the orbital period results in a very
different optical depth for the free-free absorption caused by the matter
present around these systems. 
But the dependence of $\tau_{\rm ff}$ on the square inverse of the radio frequency implies that, at higher frequencies, the effect of the surrounding matter will be less important.
For instance, applying Eq. \ref{eq:tauff} to one of the tighter orbits AMXPs, XTE~J0929--314, at 1.4 GHz, at 6.4 GHz and at 8.5 GHz, with parameters
$\rm \dot{m}_{-10}\sim 2.9$, m$_1\sim 1.4$ M$_{\odot}$, m$_2\sim$ 0.02
M$_{\odot},$ X = 0.7, Y = 0.3, $\gamma \sim$ 1, T$_4 \sim 1,$
P$_{\rm h}=0.73$ (Galloway et al. 2002) leads to $\tau_{\rm ff}$(1.4 GHz)
$\approx$ 5, $\tau_{\rm ff}$(6.5 GHz) $\approx$ 0.2, $\tau_{\rm ff}$(8.5 GHz)
$\approx$ 0.1. Adopting these two observing frequencies, the
problem of the absorption of the radiation is totally
overcome. Prompted by these considerations, we have undertaken a programme to search for millisecond pulsations in NS-SXT 
XTE~J0929--314 at 6.5 and 8.5 GHz.

The observations and the method of data analysis are presented in
Sec. \ref{section2}, and the results are reported in \S \ref{section3}
and discussed in \S \ref{section4}.

\section{Observations and data analysis}\label{section2}

\begin{table}[tt]
\begin{center}
\caption{Orbital and spin parameters for XTE~J0929--314 (Di Salvo et al. in prep.; see also Di Salvo et al. 2008b). 
The errors on the last quoted digit(s) are intended to be at $1 \sigma$ and are reported in parentheses.}
\begin{tabular}{lc}
\hline
\hline
\textbf{Parameter}        & \textbf{Value} \\
\hline
RAJ & 09$^{\rm h}$ 29$^{\rm m}$ 20$^{\rm s}$.19 \\  
DECJ & $-$31$^{\circ}$ 23$^{\prime}$ 3$^{\prime\prime}$.2 \\
Orbital period, P$_{\rm orb}$ (s) & 2614.748(3) \\
Projected semi-major axis, a$\cdot \sin $ i (lt-ms) & 5.988(10)  \\
Eccentricity, e & $<$ 0.007 \\
Spin period$^{\rm a}$, P$_{\rm S}$ (s) & 0.0054023317856(4) \\
Mean spin period  derivative, $\rm \dot{P}_{\rm S}$ (s $\cdot$ s$^{-1}$) & 1.63(12) $\times \ 10^{-18}$\\
Ascending node passage, T$_0$ (MJD) & 52405.48676(1)\\
\hline
\multicolumn{2}{l}{$^{\rm a}$ The value reported is referred at epoch 52396.5 MJD.}\\
\end{tabular}
\end{center}
\label{tab:parJ0929}
\end{table}

Radio observations of the millisecond pulsar XTE~J0929--314 were
made with the Parkes 64 m radio telescope.  Three data series were
taken on 2003 December 19-21 with a bandwidth of 576 MHz. The first was at
a central radio frequency of 6410.5 MHz, the other two at a central
radio frequency of 8453.5 MHz.

The collected signal for each polarization was split into 192 3 MHz
channels using an analogue filterbank, in order to minimize the pulse
broadening caused by dispersion in the interstellar medium (ISM). The
outputs from each channel were one bit digitized every 100 $\mu$s.
The resulting time series were stored on Digital Linear Tapes (DLT)
for off-line analysis. The observations lasted 7.5 hr each,
corresponding to $2^{28}$ samples. We then rebinned the data to obtain
$2^{27}$ samples in order to reduce the computational time; therefore,
the effective time resolution of the analysed data is 200 $\mu$s.
 
The data analysis methodology was chosen on the basis of the
precise knowledge of the orbital and spin parameters for XTE
J0929--314, obtained from the X observation. The original ephemerides
were presented by Galloway et al. (2002\nocite{gcmr02}), while we used
those reported in Table \ref{tab:parJ0929}, which have been recently
refined by Di Salvo et al. (in prep.; see also Di Salvo et al.
2008b\nocite{dbr+08b}).

Off-line analysis was made by means of a software suite that, in the
first stage, aimed to reduce the dispersion effects on the signal due
to the interstellar medium. The data series were dedispersed according
to 72 trial dispersion-measure (DM) values ranging from 5.51 to 396.74
pc$\cdot$ cm$^{-3}$ for the data series at 6410.5 MHz and 33 trial DM
values ranging from 12.64 to 416.99 pc$\cdot$ cm$^{-3}$ for the data
series at 8453.5  MHz. The maximum DM was chosen to be $\sim$ 4
times the nominal DM value obtained using either the Taylor \& Cordes
(1993\nocite{tc93}) or the Cordes \& Lazio (2001\nocite{cl01}) models
for the distribution of free electrons in the ISM, in order to account
for the errors in the estimated distance of the source, for the uncertainties in the ISM model, and for the
presence of local matter surrounding the system, hence increasing the
local DM (estimated to be at most $\sim 100$ pc/cm$^3$ for this source; Burgay et al. 2003).

The successive step is to deorbit and barycentre the time series,
i.e. to eliminate the effects of the orbital motion of the NS in the
binary system and the earth in the solar system. This was obtained by
resampling the time series in order to mimick time series collected
from a telescope located in the barycentre of the solar system and
looking at the source as if it were located at the barycentre of the
pulsar orbit.

In doing this exercise it is important to estimate the effect on the
putative pulsar signal due to the uncertainties in the adopted
ephemeris for the system. Therefore we simulated dedispersed time
series containing a periodic signal with the spin and orbital
characteristics of XTE~J0929--314, as derived by Di Salvo et al. (Table
\ref{tab:parJ0929}). We then deorbited and barycentred these
time series using the upper and the lower limits of the parameters in
Table \ref{tab:parJ0929}, adopting 1$\sigma$ errors.

It turned out that even a variation in all the parameters (but the
orbital period, see later) at $1\sigma$ level does not affect the
detectability of the pulsations, since that produced a maximum
broadening of the pulse much smaller than 0.1 in pulsar phase.  The
only exception is for the orbital period: 1$\sigma$ variation in
P$_{\rm orb}$, propagated over the $\sim$ 19500 orbits occurred between
X-ray and radio observations, produces a broadening of 0.4 in pulsar
phase for our observation. Therefore, to obtain a maximum
broadening of the pulse of 0.1 in phase, we corrected the time
series using 8 trial values of P$_{\rm orb}$ (4 above and 4 below the
nominal value), covering the 1 $\sigma$ uncertainty range.

The third step in the procedure was to fold the (already deorbited and
barycentred) time series according to the rotational parameters
reported in Table \ref{tab:parJ0929}. The trial values of the spin
period used for the folding, were chosen taking into account that
the nominal value of P$_{\rm S}$ dates from the X ray observations of May
2002, while our observations occurred 19 months later. Therefore we
explored a range of spin period $\Delta $P$_{\rm tot}$ =
P$_{\rm max}-$ P$_{\rm min}$, with P$_{\rm max}$ = P$_{\rm S}+\delta$P$_{\rm S}+\Delta$T
$(\rm \dot{P}_S+\delta\dot{P}_S)$ and P$_{\rm min}$ = P$_{\rm S} - \delta $P$_{\rm S}$,
where P$_{\rm S}$ and $\rm \dot{P}_S$ are the values reported in Table
\ref{tab:parJ0929}, $\delta$P$_{\rm S}$ and $\rm \delta \dot{P}_S$ their 1
$\sigma$ errors, and $\Delta$T is the time between the observations in
X-ray and radio bands.  As a first guess we used the outburst value of $\rm \dot{P}_S$ for this
calculation, resulting a safe upper limit for the dipolar spin-down\footnote{The
expected variation of the P$_{\rm S}$ during the radio observation is
negligible and hence does not affect a correct folding.}, as shown
below. For a safer investigation, we also checked on spin period values lower
than our previous estimate P$_{\rm min}$ and for the double of the nominal
value, 2$\times$P$_{\rm S}$.

Since the spin period derivative measured in outburst is usually
different from that in the quiescent phase, we checked the
plausibility of the adopted trial period interval through an estimate
of the surface magnetic field B$_{\rm S}$.  This estimate can be obtained
evaluating the magnetic torque acting on the NS through the formula
(Rappaport et al.  2004\nocite{rfs04}; Di Salvo et al. in prep.):
\begin{equation}\label{eq:torque}
\rm \tau (t) =\dot{M}(t)\sqrt{\rm GMR_{CO}}-\frac{\mu^{2}}{9R^{3}_{CO}}
\end{equation} 
where $\rm \dot{M}$ is the mass accretion rate, M the NS mass, $\mu$
the magnetic dipole moment, R$_{\rm CO}$ the corotation radius (at
which the disc matter in Keplerian motion rotates with the same
angular speed of the NS) in cgs units. Since this AMXP is observed to
spin down, the contribution to the spin up from the accretion (first
additive term in \ref{eq:torque}) is neglected.  The torque can also
be written:
\begin{equation}
\rm \tau (t)=2\pi I \dot{\nu}_S(t)
\end{equation} 
where I is the NS moment of inertia and $\dot{\nu}_{\rm S}$ the spin
frequency derivative. Since $\dot{\nu}_{\rm S}$ and R$_{\rm CO}$ are known and,
assuming I~$ = 10^{45}$g$\cdot$cm$^2$, we can obtain the value of the
magnetic moment and, in turn, an estimate of B$_{\rm S}$. 
Then, through the relation
\begin{equation}
\rm B=3.2 \times 10^{19}\sqrt{\rm P\ \dot{P}_{ dip}}\hspace{0.5cm}Gauss,
\end{equation} 
we can obtain an estimate of the dipolar spin-down $\rm \dot{P}_{dip}$ during observation (in quiescence),
assuming the spin down is governed by magnetodipole braking.  It
leads to B$_{\rm S}$ = 6.6 $\cdot$ 10$^{8}$ Gauss and $\rm \dot{P}_{dip} < \dot{P}_S$ and then we can 
conclude that the adopted $\Delta$P$_{\rm tot}$ is safely large. 
Therefore, since the pulse broadening for an error equal to 1/20 of the $\Delta$P interval is less than 0.1 in phase over our observations, we selected 40 trial values of the period, 20 to the right and 20 to the 
left of the nominal value to cover the $\Delta $P$_{\rm tot}$ interval calculated above.

A further check of the validity of the derived B$_{\rm S}$ (hence of the adopted $\Delta$P$_{\rm tot}$) can be derived from optical 
observations (Monelli et al. 2005\nocite{mfb+05}; D'Avanzo et al. 2008\nocite{dcc+08}). In fact, the optical counterpart of the 
XTE~J0929--314 companion observed in December 2003 (very close to our radio observations) with the VLT has a luminosity one order 
of magnitude higher than expected. This luminosity excess can be interpreted as the luminosity L$_{\rm PSR}$ isotropically irradiated
by the rotating magneto-dipole, intercepted, and reprocessed by the companion star, as observed e.g. by Burderi et al. 
(2003\nocite{bdd+03}) and Campana et al. (2004) for SAX J1808.4--3658, and by D'Avanzo et al. (2007) for IGR J00291+5934.

The excess luminosity can be written as the fraction f = f$_{\rm C}$ + f$_{\rm D}$ of dipolar radiation intercepted by the companion 
star and the disk (see Burderi et al. 2003):

\begin{equation}
\rm L_{exc} = f \cdot L_{PSR} = f \cdot \frac{2\mu^2 \omega^4}{3c^3} = f \cdot 3.85 \times 10^{35} P^{-4}_{-3}\ \mu_{26}^2 \hspace{0.2cm} erg \cdot s^{-1}
\end{equation} 
where $\omega$ is the rotational frequency of the NS and  f$_{\rm C}$ can be written as f$_{\rm C}$ = 2$\pi$a$^2$(1 $-$ $\cos \theta$)/(4$\pi$a$^2$), where a is the the orbital separation and $\theta$ the angle subtended by the companion star as seen from the central source. If the companion star fills its Roche lobe, this can be written as $\sin \theta$ = R$_{\rm L2}$/a, where R$_{\rm L2}$ is the Roche lobe radius of the secondary and R$_{\rm L2}$/a = 0.49 q$^{2/3}$/[0.6 q$^{2/3}$ + ln (1 + q$^{1/3}$)] (Eggleton 1983\nocite{egg83}). Assuming a mass ratio of q = m$_2$/m$_2$ = 0.02/1.4 and $\sin$ i = 1, we obtain f$_{\rm C}$ = 0.003; while f$_{\rm D}$ = 0.012 is evaluated adopting a standard Shakura-Sunyaev disk model (Shakura \& Sunyaev 1973\nocite{ss73a}) with $\rm \dot{m}_{-10}$ = 2.9, and is given by the projected area of the disk as seen by the central source, 2$\pi$R$\times$2H(R) (where R is the disk outer radius and H(R) the disk semi-thickness at R), divided by the total area, 4$\pi$R$^2$.

Adopting the value of L$_{\rm exc}$ $\simeq$ 3.33 $\times$ 10$^{32}$ erg $\cdot$ s$^{-1}$ reported by Monelli et al. (2005) and f $\simeq$ 1.15 $\times$ 10$^{-2}$, we can thus derive B$_{\rm S}$ $\simeq$ 7 $\cdot$ 10$^{8}$ Gauss, in good agreement with the value estimated above.

\section{Results}\label{section3}

\begin{figure}[tt]
\begin{center}
\includegraphics[width=8.8cm]{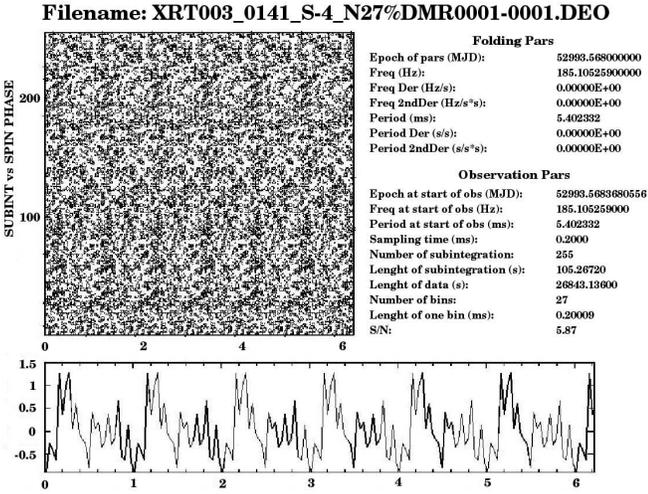}
\end{center}
\caption{Example of a plot resulting from the procedure described in
the text. The bottom diagram shows six integrated pulse profiles, while
the grayscale on the left represents the signal in the 255
subintegrations in which the observation has been subdivided. On the
right, the parameters used for the folding are shown, along with the
parameters of the observation.}
\label{fig:candidate}
\end{figure}

Spanning the searched ranges of P$_{\rm orb}$, P$_{\rm S}$ and DM we have
obtained 50922 plots, reporting the result from the folding of the
deorbited and barycentred time series. 
An example of these plots (obtained from the observation at 6.5 GHz) is shown in
Fig. \ref{fig:candidate}. The bottom diagram shows the integrated
pulse profile (6 phases), while the grayscale on the left represents
the signal in the 255 subintegrations in which the observation has
been subdivided. A good candidate should display a roughly linear
trend in the grayscale and a high signal-to-noise ratio pulse
profile. On the right, the parameters used for the folding of this
candidate are shown, along with the parameters of the observation.

\begin{figure}[tt]
\begin{center}
\includegraphics[width=5.8cm,angle=-90]{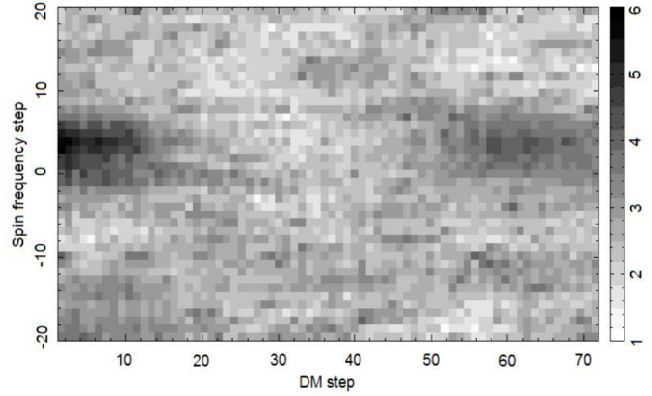}
\end{center}
\caption{Signal-to-noise ratio in function of the spin frequency (20
step above and 20 below the nominal value, corresponding to 0) and the
dispersion measure (72 steps corresponding to values interval from 5.51
to 396.74 cm$^{-3}\cdot$ pc) at P$_{\rm orb}$ = 2614.74575 s.
The highest point corresponds to the best S/N, $\approx$ 6, reported
in Fig. \ref{fig:candidate}.}
\label{fig:snmax}
\end{figure}

A useful diagnostic tool that can be adopted to further evaluate the
plausibility of a suspect is shown in Fig. \ref{fig:snmax}, where a
grayscale of the strength of the signal (with darker points at
higher S/N), plotted as a function of dispersion measure DM and spin frequency
$\nu_{\rm S}$, should define a clear peak around the correct parameters of the
putative pulsar. The same kind of plot can be created at a constant DM
with $\nu_{\rm S}$ and P$_{\rm orb}$ on the axes or at a constant $\nu_{\rm S}$,
tracing the S/N trend at varying DMs and P$_{\rm orb}$s.

The plots in Figs. \ref{fig:candidate} and \ref{fig:snmax} refer to the suspect with the highest S/N found in our 
search. Its folding parameters are listed in Table \ref{tab:parcandidate}. The peak in the profile has a 3.4 $\sigma$ significance that, 
over the 26568 trial foldings performed on the 6.5 GHz dataset (72 DMs $\times$ 9 P$_{\rm orb}$s $\times$ 41 P$_{\rm S}$s), has a 
probability of not being randomly generated by noise of $\sim$ 10$^{-6}$. We also note that, in Fig. \ref{fig:snmax} 
the decreasing S/N trend is not particularly defined, although a maximum is present. Also, that the DM that maximises the S/N is 
close to zero (w.r.t an expected value of $\gsim$ 100 pc/cm$^3$) weakens the credibility of the signal. Finally, this signal suspect is 
not confirmed in the observations at higher frequency (8.5 GHz), elaborated with the same parameters.
We can conclude that no radio pulsation with the expected
periodicity has been found in the source XTE~J0929--314 during its
quiescent phase.

\begin{table}
\caption{Parameters for the pulsar candidate with the highest S/N shown in Fig. \ref{fig:candidate}.}
\label{tab:parcandidate}
\centering                          
\begin{tabular}{l c}        
\hline\hline                 
Signal to noise ratio, S/N & 5.87\\
Dispersion Measure, DM (pc cm$^{-3})$ & 5.51 \\
Orbital period, P$_{\rm orb}$ (s) & 2614.74575 \\
Projected semi-major axis, a$\cdot \sin$ i (lt-ms) & 6.290(9) \\
Spin frequency, $\nu_{\rm S}$ (Hz) & 185.105259 \\
Ascending node passage, T$_0$ (MJD) & 52405.48676(1) \\
\hline                        
\hline                                   
\end{tabular}
\end{table}

\subsection{Flux density upper limits}
\begin{figure}[tt]
\begin{center}
\includegraphics[width=8.8cm]{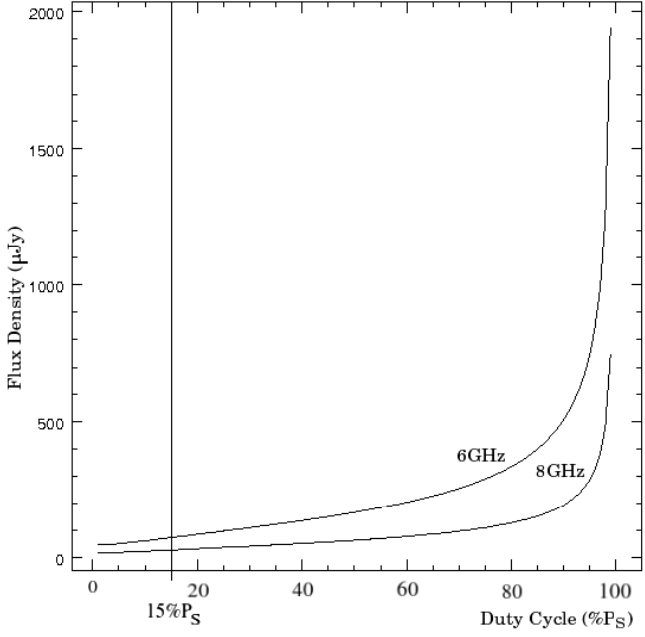}
\end{center}
\caption{Flux density upper limits in $\mu$Jy for XTE~J0929--314
obtained with Eq. \ref{eq:sensi} for increasing values of the
duty cycle.}
\label{fig:fluxduty}
\end{figure}

A rough estimate of the flux density for a pulsar of period P is
(e.g. Manchester et al. 1996\nocite{mld+96})
\begin{equation}\label{eq:sensi}
\rm S=\epsilon \ n_{\sigma} \frac{T_{sys}+T_{sky}}{G\sqrt{\rm N_p\ \Delta t \ \Delta \nu_{MHz}}}\sqrt{\rm \frac{W_e}{P-W_e}} \hspace{0.5cm} mJy,
\end{equation}
where n$_{\sigma}$ = 6.0 is the threshold (in unit of sigma) for having a statistically significant signal, given the number of performed foldings;
T$_{\rm sys}$ = 50 K the system noise temperature for the
observations at $\sim$ 6 GHz and T$_{\rm sys}$ = 25 K for those at $\sim$
8 GHz (see Parkes website: http://www.parkes.atnf.csiro.au/observing/documentation);
T$_{\rm sky}$ is the sky temperature in Kelvin, calculated from that at
$\nu$ = 408 MHz and considering a scaling with the frequency as
$\nu^{-2.7}$; G is the gain of the radio telescope (in K $\cdot$
Jy$^{-1}$), with 0.46 for the observations at $\sim$ 6 GHz and 0.59
for those at $\sim$ 8 GHz, (see Parkes website); $\Delta $t is the integration time
in seconds, N$_{\rm p}$ the number of polarizations (here 2), and $\Delta
\nu_{\rm MHz}$ = 576 MHz is the bandwidth in MHz. The term $\epsilon \sim$
1.4 is a factor accounting for the sensitivity reduction due to
digitisation and other losses. The term W$_{\rm e}$ is the effective width
of the pulse:
\begin{equation}\label{eq:allargamento}
\rm W_e = \sqrt{\rm W^2 + (\beta \delta t)^2 + \delta t^2_{DM} + \delta t^2_{scatt}}.
\end{equation}
Its value depends on the sampling time ($\delta$t = 200 $\mu$s in this
case), on the technical characteristics of the receiver ($\beta$ = 2),
on the broadening of the pulse introduced by both the dispersion of
the signal in each channel ($\delta $t$_{\rm DM}\sim 4\div 9\times
10^{-8}\times$ DM sec, depending on the frequency), on the scattering
induced by inhomogeneities in the ISM ($\delta$t$_{\rm scatt} \sim
10^{-9}$ sec), and on the intrinsic pulse width W.  The dependence of the
minimum flux density achievable from the duty cycle, W/P, is shown in
Fig. \ref{fig:fluxduty}.
A duty cycle of 15\% of the spin period is assumed and, with this
value, the flux density upper limits for XTE~J0929--314 are:
S$^{6.5}_{\rm MAX}$ = 68 $\mu$Jy at 6410.5 MHz and S$^{8.5}_{\rm MAX}$ = 26
$\mu$Jy at 8453.5 MHz.

\section{Discussion}\label{section4}

In the hypothesis that a radio pulsar was on during our observations, we examine several possible
explanations for the lack of the detection of radio pulsation from our source.

\subsection{The luminosity}
\begin{figure}[tt]
\begin{center}
\includegraphics[width=9cm]{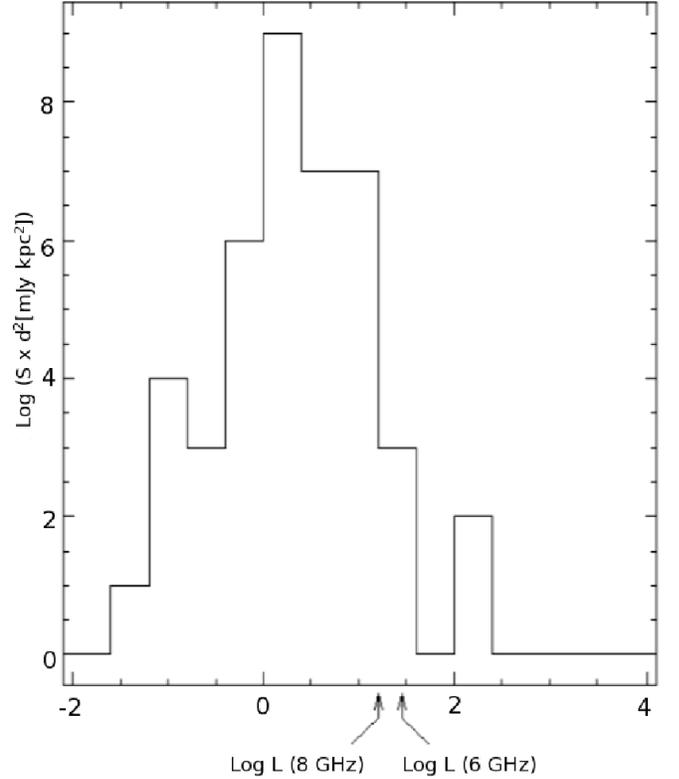}
\end{center}
\caption{Pseudoluminosity distribution of a sample of 42 field
MSPs. The arrows indicate the upper limits of the pseudoluminosity
scaled at 1.4 GHz for XTE~J0929--314, derived on the basis of the
minimum flux detectable and assuming a distance of 6 kpc (Galloway et
al.  2002).}
\label{fig:lumi}
\end{figure}

In Fig. \ref{fig:lumi} the pseudoluminosity L = S$\cdot$d$^2$ in
mJy$\cdot$kpc$^2$ (where S is the measured flux density and d the
distance of the source) distribution at 1.4 GHz for a sample of 42
galactic field MSPs is shown\footnote{These values are derived from
the ATNF catalogue -
http://www.atnf.csiro.au/research/pulsar/psrcat/; Manchester et al. 2005\nocite{mht+05}}.
For XTE~J0929--314 the upper limits on the pseudoluminosity has been
calculated assuming a distance of 6 kpc (Galloway et al. 2002) and our
flux density upper limits at 6.4 and 8.5 GHz scaled to 1.4 GHz
assuming a spectral index for MSPs of 1.7 (Kramer et al
1998\nocite{kxl+98}). As a result, about 90\% of the known MSPs are below the
pseudoluminosity upper limits of XTE~J0929--314.

This suggests that XTE~J0929--314 in quiescence might not be a
very bright MSP and that a deeper search should be carried out to
sample lower values of the flux density. However, given the distance
of XTE~J0929--314 (significantly higher than the typical distance of
the sample of known MSPs), only next-generation instruments, like the
Square Kilometer Array, will be able to discover pulsed radio emission
from this source at the high frequencies adopted in this work, if its
luminosity is at the faintest end of the known distribution.

\subsection{The beaming factor}
The emission from a pulsar is strongly anisotropic. This means that it
does not irradiate in the same way in all directions, but only on
one usually narrow portion of sky, which can be quantified through the
so-called beaming factor. Therefore, the lack of a detection of a radio signal could be due to
unfavourable geometry of the radio emission with respect to the
observer.

The average value, f$(\alpha)$, of the fraction of the sky swept from
two conal radio beams of half width $\alpha$ is (Emmering \& Chevalier
1989\nocite{ec89}):
\begin{equation}\label{eq:falfa}
\rm f(\alpha) =  \int^{\pi /2}_{0} f(\alpha , \eta) \sin \eta \ d\eta = (1-\cos \alpha) + \left(\frac{\pi}{2}- \alpha \right)\sin \alpha
\end{equation}
where $\eta$ is the angle (supposed randomly distributed) between the
magnetic axis (aligned with the radio beams) and the rotational axis,
and f$(\alpha,\eta)=\cos$[max(0, $\eta-\alpha)]-\cos$[min(0, $\eta+\alpha )]$.

Considering a 10\% $\div$ 30\% interval of width of the pulse, it
follows that $9^{\circ} \leq \alpha \leq 27^{\circ}$ and, therefore, using
Eq. \ref{eq:falfa}, 0.23 $\leq$ f($\alpha)\leq$ 0.61. In particular,
assuming 15\% typical value of width of the impulse, the probability
that the radio emission cone does not intersect our line of sight is
66\%, with f($\alpha) \approx$ 0.34.

\section{Conclusions}\label{section5}
The aim of this work was to search for radio pulsations from the AMXP
XTE~J0929--314. The detection of radio signals in the quiescent phase
of this kind of transient systems would be ultimate proof of the
recycling model, unambiguously establishing that the AMXPs are the
progenitors of radio MSP. No radio pulsation has been detected in the
analysed data down to a limit of 68 $\mu$Jy at 6.4 GHz and 26 $\mu$Jy
at 8.5 GHz. Assuming that a radio pulsar was on during the observation,
the free-free absorption cannot be responsible for the lack of a
detection, given the relatively high radio frequencies. The
beaming factor is a viable explanation for that, since the probability
of any unfavourable geometry of radio emission with respect to the
observer is $\sim 60\div 70 \%$.  However, the most likely reason for a
negative result is that the source has a luminosity lower than our
limits. In fact, more than 90\% of the known MSPs show a luminosity
lower than the upper limits that we have derived for XTE~J0929--314.

To get round to these problems, one can make observations at an intermediate frequency between 6.5 and 1.4 GHz. 
For a maximum mass transfer rate in quiescence equal to the outburst value, we can estimate that the lower limit in frequency for 
obtaining $\tau_{\rm ff}\sim 1$ is $\sim$ 3 
GHz. Calculating S$_{\rm MAX}$ for an observation with all the same parameters (adopted in the present work)  
but the frequency (now set at 3 GHz), we could sample more than a half of the known MSPs luminosity distribution.

\begin{acknowledgements}
A.P. and M.B. acknowledge the financial support to this research provided by the Ministero dell'Istruzione, dell'Universit\`a e della 
Ricerca (MIUR) under the national programme PRIN052005024090\_002.

\end{acknowledgements}

\end{document}